\documentclass[tcad, a4paper]{IEEEtran}

\usepackage{amsmath,amsfonts}
\usepackage{algorithmic}
\usepackage{array}
\usepackage[caption=false,font=normalsize,labelfont=sf,textfont=sf]{subfig}
\usepackage{textcomp}
\usepackage{stfloats}
\usepackage{url}
\usepackage{verbatim}
\usepackage{graphicx}
\usepackage{hyperref}
\usepackage{booktabs}
\usepackage{amsmath}
\usepackage[style=ieee, bibencoding=utf8]{biblatex} 
  
\hypersetup{
    colorlinks=true,       
    linkcolor=blue,        
    citecolor=blue,       
    filecolor=blue,     
    urlcolor=blue          
}
\usepackage{pifont} 
\usepackage{bbding}

\usepackage{balance}

\begin{document}
\title{E2ATST: A Temporal-Spatial Optimized Energy-Efficient Architecture for Training  Spiking Transformer}
\author{Yunhao Ma, Yanyu Lin, Mingjing Li, Puli Quan, Chenlin Zhou, Wenyue Zhang, Zhiwei Zhong, Wanyi Jia, Xueke Zhu, Qingyan Meng, Huihui Zhou and Fengwei An, \textit{Member, IEEE} 
\thanks{This work is supported by the key project of the Pengcheng Laboratory (2433811500001) and the Shenzhen Science and Technology Program Shenzhen Science and Technology Program (Grant 
No. KJZD20230923113300002, JCYJ20241206180301001, 
KJZD20230923113300002, JCYJ20241206180301003, and
KQTD20200820113051096). Corresponding authors: Huihui Zhou and Fengwei An.

Yunhao Ma and Wenyue Zhang are with the Southern University of Science and Technology, Shenzhen 518055, and also with Pengcheng Laboratory, Shenzhen 518055, China. (e-mail: 12331325@mail.sustech.edu.cn; 12231163@mail.sustech.edu.cn)

Yanyu Lin, Mingjing Li, Puli Quan, Zhiwei Zhong, Xueke Zhu, Qingyan Meng and Huihui Zhou are with the Pengcheng Laboratory, Shenzhen 518055, China. (e-mail: linyy01@pcl.ac.cn; limj@pcl.ac.cn; quanpl@pcl.ac.cn; zhongzhw01@pcl.ac.cn; zhuxk@pcl.ac.cn; mengqy@pcl.ac.cn; zhouhh@pcl.ac.cn)

Chenlin Zhou is with the Peking University, Shenzhen 518000, China and also with PengCheng Laboratory, Shenzhen 518055, China. (e-mail: chenlinzhou25@stu.pku.edu.cn)

Wanyi Jia is with the Shenzhen Institutes of Advanced Technology, the Chinese Academy of Sciences, and Pengcheng Laboratory Shenzhen 518055 and with the University of Chinese Academy of Sciences, Beijing 101408, China. (e-mail: jiawanyi23@mails.ucas.ac.cn)

Fengwei An is with the Southern University of Science and Technology, Shenzhen 518055, China. (e-mail: anfw@sustech.edu.cn)
}}

\maketitle

\begin{abstract}
Neuromorphic computing and Spiking Neural Networks (SNNs) have emerged as energy-efficient alternatives to traditional artificial intelligence architectures. In parallel, Transformer models deliver state-of-the-art performance across domains but rely on dense computations and frequent memory access. Spiking Transformers, hybrid models integrating SNN sparsity with Transformer attention mechanisms, bridge this gap, yet their development is hindered by critical challenges: mainstream accelerators lack optimizations for their training workflows, and no systematic framework exists to evaluate training-related latency, energy, or dataflow tradeoffs, slowing hardware innovation.
To address these, this work presents the training architecture tailored for Spiking Transformers. Unlike existing designs mainly focus on general models or with inference function only, this architecture uniquely handles temporal signals, spatial spike signals and membrane potential signals during training, enabling end-to-end Spiking Transformer simulation of latency and energy. A dedicated temporal-spatial simulation framework further quantifies dataflow optimizations, guiding hardware tradeoff decisions and further the real hardware implementations. With its optimal dataflow found in our study, the architecture achieves 2.36 TFLOPS/W energy efficiency, outperforming State-Of-The-Art(SOTA) accelerators while retaining FP16 precision and full training support, balancing accuracy and efficiency, and supporting the learning ability of neuromorphic architectures.
This work validates the capability of low-energy Spiking Transformer training and provides a foundational methodology for hardware design, accelerating the deployment of hybrid neuromorphic-Transformer systems in energy-constrained environments.
\end{abstract}


\section{Introduction}
\IEEEPARstart{N}{euromorphic} computing, inspired by the structural and functional principles of biological neural systems, has attracted increasing attention as a potential solution to the energy bottlenecks of traditional Artificial Intelligence (AI) architectures \cite{neurpmorphoic}, \cite{TCASAI2}. Spiking Neural Networks (SNNs) are at the core of this paradigm \cite{SNN}. Unlike conventional artificial neural networks, which rely on continuous-valued signals and dense synchronous computation, SNNs use sparse and asynchronous spike-based signaling to transmit and process information. This event-driven nature allows computation to occur only when necessary, substantially reducing switching activity and dynamic power consumption \cite{10848017}, \cite{jetcas1}, \cite{TCASAI3}. As a result, SNNs are inherently well-suited for energy-constrained applications such as edge computing, autonomous sensing, and always-on devices. Recent advances in SNN hardware and training algorithms have further enhanced their potential for deployment in real-world intelligent systems.

On the other hand, the Transformer \cite{Vaswani2017AttentionIA} models have become the backbone of modern artificial intelligence due to their superior performance across a wide range of domains, including natural language processing, computer vision, and reinforcement learning \cite{zeng2024flightllm}, \cite{COSA}, \cite{TCASAI1} and \cite{dong-isscc}. However, their success comes at a high computational and energy cost. Transformer architectures typically rely on large-scale dense matrix multiplications, attention heads, and feedforward networks, all of which require frequent memory access and intensive floating-point computation \cite{ViA}. This results in significant power draw and heat generation, particularly during training or large-scale inference. Although high-performance accelerators such as Graphics Processing Unit (GPU) and Tensor Processing Unit (TPU) can deliver impressive throughput, they are often limited in energy efficiency, making them less suitable for real-time or embedded applications \cite{Transformer_train}.

To bridge the gap between energy efficiency and representational power, emerging works are exploring hybrid models that integrate SNN dynamics with Transformer-based components. These hybrid architectures aim to leverage the sparsity and temporal coding of SNNs while preserving the ability of Transformers to capture complex dependencies through attention mechanisms \cite{zhou2023spikformer}, \cite{Zhou2024QKFormerHS}, \cite{yao2023spikedriven}, and our previous work \cite{zhou2023spikingformer}. Such designs offer a path toward high-performance yet energy-efficient computation by reducing redundant activity and minimizing the need for high-precision floating-point operations. As intelligent computing continues to scale, the energy gap between brain-inspired and Transformer-based systems becomes more pronounced. However, mainstream computing platforms such as GPUs lack specialized optimizations for training these spike-driven Transformer models, making it difficult to meet their low-power advantages \cite{H2Learn}. Moreover, as the number of parameters in models grows significantly, the amount of data generated during training increases exponentially, especially those associated with time steps, leading to intensive memory access and storage demands. This results in extremely high energy consumption during computation.

Recent advances in Spiking Transformer accelerators have enabled highly parallel and energy-efficient implementations \cite{apccas} and \cite{Li2023EfficientDS}, mainly for only inference. However, whether the architecture design, data flow optimization, and energy efficiency have been optimized remains a key issue needed to be solved urgently. Despite advances in hardware‑accelerated inference for conventional deep neural networks (DNNs), including mature energy evaluation frameworks and tools such as LLMCompass \cite{LLMCompass}, Timeloop \cite{timeloop}, SCALE-simv3 \cite{raj2025scale} and the simulation research EOCAS \cite{Ma2025EnergyOrientedCA} and SimST [need cite] for SNN training and inferencing, there is no systematic evaluation methodology for hardware consumption in training Spiking Transformer.

To address these challenges, this work presents the first training acceleration architecture (to the best of our knowledge) tailored for SpikingFormer-style Spiking Transformers. The proposed design comprehensively considers both temporal and spatial spike-related signals during the training process, enabling an overall architectural simulation of latency and energy consumption. This provides valuable insights for future hardware implementations of Spiking Transformer training or fine-tuning, proposing standardized guidance for the design of SNN Transformer training accelerator chips and offering forward-looking references for hardware-oriented design in low-power SNN network architecture development. Specifically, the key contributions of this study are as follows:

1. First Specialized Training Architecture for Spiking Transformers. This work is the first to propose a training acceleration architecture tailored for SpikingFormer-style Spiking Transformers. Unlike existing accelerators that focus on DNN inference, pure SNNs, or general Transformer inference, it uniquely integrates the handling of temporal and spatial spike signals in training, directly addressing the hybrid characteristics of Spiking Transformers that traditional mainstream accelerators ignore. This specialization fills the gap of dedicated hardware for training such hybrid models.
 
2. Systematic Temporal-Spatial Simulation for Training Workflows. A comprehensive architectural training workflow and simulation framework are developed to model energy consumption during training. Unlike existing tools that lack support for hybrid Spike-Attention training, this framework enables quantitative evaluation of dataflow optimizations and hardware tradeoffs, providing a systematic methodology for Spiking Transformer training hardware design.
 
3. Balanced Design for Efficiency and Training Compatibility. The architecture achieves a unique balance between energy efficiency and training capability. By leveraging spike-driven sparsity and optimizing memory hierarchy for temporal-spatial spike signals, it reaches 2.36 TFLOPS/W energy efficiency, outperforming other State-Of-The-Art (SOTA) works while retaining FP16 precision and full training support, making it suitable for Spiking Transformer deployment.

\section{Representative Spiking Transformer Model: SpikingFormer}
\subsection{The LIF Spiking Neuron Model}
Spike neuron is the fundamental unit of SNNs; we choose the Leaky Integrate-and-Fire (LIF) \cite{LIF} model as the spike neuron in our work. The dynamics can be formulated as follows:
\begin{gather}
H[t]=V[t-1]+\frac{1}{\tau}\left(X[t]-\left(V[t-1]-V_{\text {reset }}\right)\right),\\
S[t]=\Theta\left(H[t]-V_{t h}\right) , \\
V[t]=H[t](1-S[t])+V_{\text {reset }} S[t]
\end{gather}
where $\tau$ is the membrane time constant, and $X[t]$ is the input current at time step $t$. When the membrane potential $H[t]$ exceeds the firing threshold $V_{th}$, the spike neuron will trigger a spike $S[t]$. $\Theta(v)$ is the Heaviside step function, which equals to 1 when $v\geq 0$ and 0 otherwise. $V[t]$ represents the membrane potential after the triggered event, which equals to $H[t]$ if no spike is generated and otherwise equals to the reset potential $V_{reset}$. 

\begin{figure*}[htbp]
    \centering
    \includegraphics[width=1\textwidth,page=1]{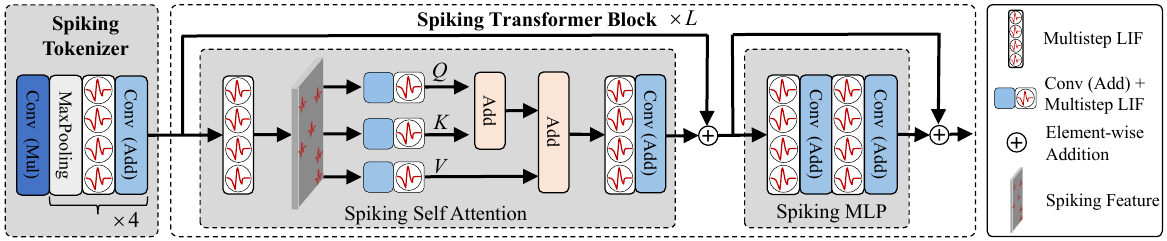}
    \caption{
    Overview of the Spiking Transformer algorithms (Spikingformer \cite{zhou2023spikingformer}), which consists of a Spiking Tokenizer, several Spiking Transformer Blocks.}
    \label{Spikingformer}
\end{figure*}

\begin{table}[htb]
\caption{Model comparison. }
\vspace{-18pt}
\begin{center}
\begin{tabular}{lccc}
\toprule
{Models} & ViT-B/16 \cite{dosovitskiy2020image}  & Spikformer \cite{zhou2023spikformer} &  Spikingformer \cite{zhou2023spikingformer} \\
\midrule
Test Size & 224$\times$224  & 224$\times$224   & 224$\times$224 \\ 
Model Size & 86M  & 66.34M  & 66.34M \\ 
Model Type &  ANN  & SNN  & SNN  \\ 
Attention &  SA   & SSA  & SSA  \\ 
Residual &  Vanilla Add    & SEW Add      & MS Add  \\ 
Spike-Driven &  \ding{55}  & \ding{55}  & \ding{51}  \\ 
OPs (G) &  17.6  & 22.09  & 12.54  \\ 
\midrule
Energy (mJ) & 80.9   & 32.07  & 13.68  \\ 
ImageNet-1K  &  77.91  & 74.81 &  75.85 (77.64 \cite{zhou2023enhancing}) \\   
\bottomrule
\end{tabular}
\end{center}
\vspace{-0.5cm}
\label{complexity}
\end{table}

\subsection{Overall Forward Computing}
Our proposed Spikingformer contains a  Spiking Tokenizer (ST), several Spiking Transformer Blocks, and a Classification Head. Given a 2D image sequence $\textbf{I} \in \mathbb{R}^{T \times C \times H \times W}$ (Note that $C \text{=} 3$ in static datasets like ImageNet 2012, $C \text{=} 2$ in neuromorphic datasets like DVS-Gesture), we use the Spiking Tokenizer block for downsampling and patch embedding, where the inputs can be projected as spike-form patches $\textbf{X} \in \mathbb{R}^{T \times N \times D}$. Obviously, the first layer of Spiking Tokenizer also plays a spike encoder role when taking static images as input. After Spiking Tokenizer, 
the spiking patches $\textbf{X}_{0}$ will pass to the $L$ Spiking Transformer Blocks.
Similar to the standard ViT encoder block, a Spiking Transformer Block contains a Spiking Self Attention (SSA) \cite{zhou2023spikformer} and a Spiking MLP block. In the last, a fully-connected-layer (FC) is used for the Classification Head. Note that we use a global average-pooling (GAP) before the fully-connected layer to reduce the parameters of FC and improve the classification capability of Spikingformer.
\begin{align}
&\textbf{X} =\operatorname{ST}(\textbf{I}), \quad \textbf{I} \in \mathbb{R}^{T \times C \times H \times W}, \textbf{X} \in \mathbb{R}^{T \times N \times D} \\
&\textbf{X}_l^{\prime}=\operatorname{SSA}\left(\textbf{X}_{l-1}\right)+\textbf{X}_{l-1},  \textbf{X}_l^{\prime} \in \mathbb{R}^{T \times N \times D},  \\
&\textbf{X}_l=\operatorname{SMLP}\left(\textbf{X}_l^{\prime}\right)+\textbf{X}_l^{\prime},  \textbf{X}_l \in \mathbb{R}^{T \times N \times D}, \\
&\textbf{Y}=\operatorname{FC}\left(\operatorname{GAP}\left(\textbf{X}_L\right)\right).  
\end{align}

The Spiking Transformer Block contains a Pre-activation Spiking Self-Attention (PSSA) block and a Spiking MLP block. PSSA retains Spiking Self-Attention (SSA)'s global modeling capabilities while making modifications to spike-driven and more generalization. Thus, PSSA can be seen as an important variant of SSA. Therefore, the PSSA can be formulated as follows:
\begin{gather}
\textbf{X}^{\prime} = \operatorname{SN}(\textbf{X}),
\end{gather}
\begin{equation}
\left\{
\begin{aligned}
\textbf{Q} &= \operatorname{SN}_{Q}(\operatorname{Conv1DBN}_{Q}(\textbf{X}^{\prime})), \\
\textbf{K} &= \operatorname{SN}_{K}(\operatorname{Conv1DBN}_{K}(\textbf{X}^{\prime})), \\
\textbf{V} &= \operatorname{SN}_{V}(\operatorname{Conv1DBN}_{V}(\textbf{X}^{\prime})),
\end{aligned}
\right.
\end{equation}
\begin{gather}
\operatorname{SSA}(\textbf{Q}, \textbf{K}, \textbf{V})= \operatorname{Conv1DBN}(\operatorname{SN}(\textbf{Q} \textbf{K}^{\mathrm{T}} \textbf{V} * s)),
\end{gather}
where $\textbf{Q}, \textbf{K}, \textbf{V} \in \mathbb{R}^{T \times N \times D}$ are pure spike data (only containing 0 and 1). $s$ is the scaling factor as in \cite{zhou2023spikformer}, controlling the large value of the matrix multiplication
result.

\subsection{BPTT for SNN-based Spiking Transformer}
Due to the spatiotemporal data paths in SNN-based models, backpropagation through time (BPTT) is a good fit \cite{H2Learn}, which has been proved through its accuracy demonstration \cite{jetcas1}, \cite{NC} and \cite{SATA}. Following variables in Table \ref{tab:variable_descriptions}, the generation of spike-related signals are shown as follows:
\begin{equation}
\label{fppass}
\begin{cases}
U_t^l[x] = {\alpha U_{t-1}^l[x] \bigl(1 - S_{t-1}^l[x]\bigr)} + BN_t^{l}[x] \\
S_t^l[x] = {fire}\bigl(U_t^l[x] - {th}_f\bigr)=\begin{cases}
    1, & U_t^l[x] \geq {th}_f \\ 
    0, & U_t^l[x] < {th}_f 
\end{cases}\\
\nabla \tilde{S}[x] = \begin{cases}
    1, & {th}_f < U_t^l[x] <{th}_r \\
    0, & \text{other cases}
\end{cases}
\end{cases}
\end{equation}
Here $t$ stands for the time step, and $l$ is the layer and $x$ denotes the neuron. Additionally, $\alpha$ means the leakage factor. When the potential crosses a threshold ${th}_f$, the neuron fires a spike and resets its potential to zero. ${fire}(\cdot)$ is the Heaviside step function, i.e., ${fire}(x) = 1$ if $x \geq 0$; ${fire}(x) = 0$ otherwise. $BN_t^l[x]$ indicates the result of the $x$-th batch normalization at time step $t$ and layer $l$.

During the backward pass, the generation of spike-related signals can be expressed as:
\begin{equation}
\begin{cases}
\nabla S_t^l[x] = {\nabla U_{t+1}^l[x] \bigl(-\alpha U_t^l[x]\bigr)} + MM_t^{l}[x] \\
\nabla U_t^l[x] = \nabla U_{t+1}^l[x] \alpha \bigl(1 - S_t^l[x]\bigr) + \nabla S_t^l[x] {fire}'\bigl(U_t^l[x]\bigr)\\
\nabla \tilde{S}_t^{l}[x]=fire'(U_t^l[x])
\end{cases}
\end{equation}
Here, \(MM_t^l[x]\) represents the backward pass result of $x$-th matrix multiplications at time step $t$ and layer $l$.
\section{Architecture Design}
\subsection{Detailed Training Flow of Spiking Transformer}
\begin{figure*}[!htbp]
    \centering
    \includegraphics[width=1\textwidth,page=1]{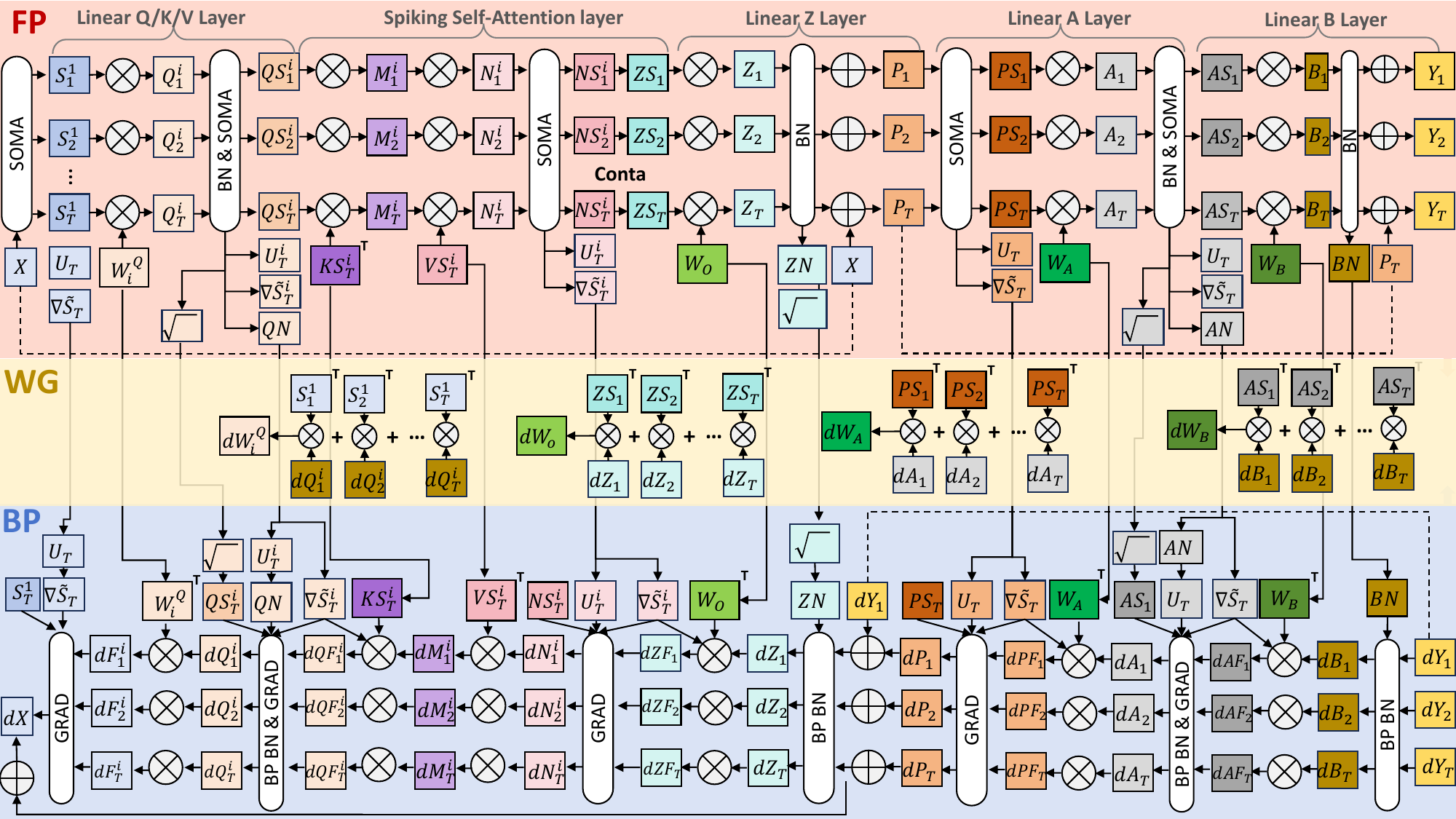}
    \caption{The detailed training flow of a representative model of the Spiking Transformer, encompassing detailed computations of FP/BP BN, SOMA, and GRAD across the three stages (FP, BP, WG) as well as the relationships between various data. (Connections within the same stage are denoted by dashed lines, while connections associated with different stages are denoted by solid lines.) The training flow is inspired from the training studies \cite{Transformer_train} and \cite{NC}.}
    \label{fig_overall_training}
\end{figure*}

\begin{table}[!t]
\centering
\caption{Variable descriptions for the related quantities.}
\renewcommand{\arraystretch}{1.3} 
\begin{tabular}{|c|c||c|c|}
\hline
\textbf{Var} & \textbf{Description} & \textbf{Var} & \textbf{Description} \\ \hline
$S$          & spike                & $U$          & membrane potential   \\ \hline
$\nabla S$   & spike gradient       & $\nabla U$   & potential gradient   \\ \hline
$\nabla \tilde{S}$ & spike gradient mask & $\nabla \tilde{U}$ & potential gradient mask \\ \hline
\end{tabular}

\label{tab:variable_descriptions}
\end{table}
Fig. \ref{fig_overall_training} illustrates the training computation workflow of SpikingFormer, a typical example of a general Spiking Transformer. Due to the involvement of time steps, there are always spike state signals emitted at multiple time points, generating corresponding membrane potentials \( U_T \), spike signals \( S_T \), and spike gradient mask signals \( \nabla \tilde{S_T} \); these signals are utilized in subsequent Backward Propagation (BP) and Weight Update based on Gradient (WG) stages, which are the three phases in BPTT learning. Unlike traditional Transformer architectures, therefore, the entire training computation process incorporates temporal dimension information and does not include softmax operations. The Conv1D mentioned in SpikingFormer can be regarded as matrix multiplication (MM).  

In the process of FP (Forward Propagation), matrix operations in the Q/K/V linear transformation layers are performed via floating-point addition (FP16), as they only involve computations between spike signals and FP16 values, which can be simplified to addition operations. Consequently, the three matrices \(Q_T^i\), \(K_T^i\), and \(V_T^i\) are converted into FP16 matrices after matrix operations. Subsequently, through batch normalization (BN) and SOMA (the spike deliver module from \cite{NC} conducting $fire$ operations in formula \ref{fppass}), they are transformed back into spike binary signals, i.e., \(QS_T^i\), where $T$ denotes the time step and $i$ represents the number of multi-heads. It should be noted that all variables with the suffix $S$ correspond to spike signals, and these signals are converted into FP16 data format during BP; thus, the $F$ in \(dQF_T^i\) in the figure indicates FP16. Next, computations in the attention mechanism layer are performed, where matrix calculations are also completed using spike addition instead of multiplication. Finally, the summed and scaled spike signals are output to the Linear Z layer for further matrix operations, followed by BN calculations and residual connections. Subsequently, linear transformations, BN operations, and corresponding residual connections in the Linear A/B layers are executed. Throughout FP, the output parameters of BN need to be retained to enable the calculation of BN loss during BP.
In the process of BP, the loss input also contains temporal dimension information. Its computational flow is generally opposite to and one-to-one corresponding with that of FP. However, for training accuracy, all BP signals are in FP16, which means matrix multiplication must support FP16 floating-point multiplication. In BN process of BP, the BN parameters corresponding to the FP part are required for computation. For each reverse linear transformation multiplication, the transposed weights of the corresponding forward linear transformation and the spike gradient mask signal \(\tilde{\nabla S}_T\) are needed. For each reverse SOMA operation (GRAD operation), the input signals must include the gradient from the subsequent BP layer, the spike signals of the corresponding FP layer, membrane potentials, and the spike gradient mask \(\nabla\tilde{ S_T}\).

In the process of WG, the total involves the sum of weights from three linear layers in both the forward and backward stages. These linear transformations can all be completed by transposing the corresponding spike results in FP and performing matrix multiplication with the results of each BP layer. These matrix multiplications are operations between spikes and FP16, so the multiplication process can also be simplified to addition operations.

\subsection{Overall Architecture of the Accelerator}
\begin{figure}[!t]
\centering
\includegraphics[width=0.5\textwidth,page=2,trim={30bp 0bp 120bp 0bp}, clip]{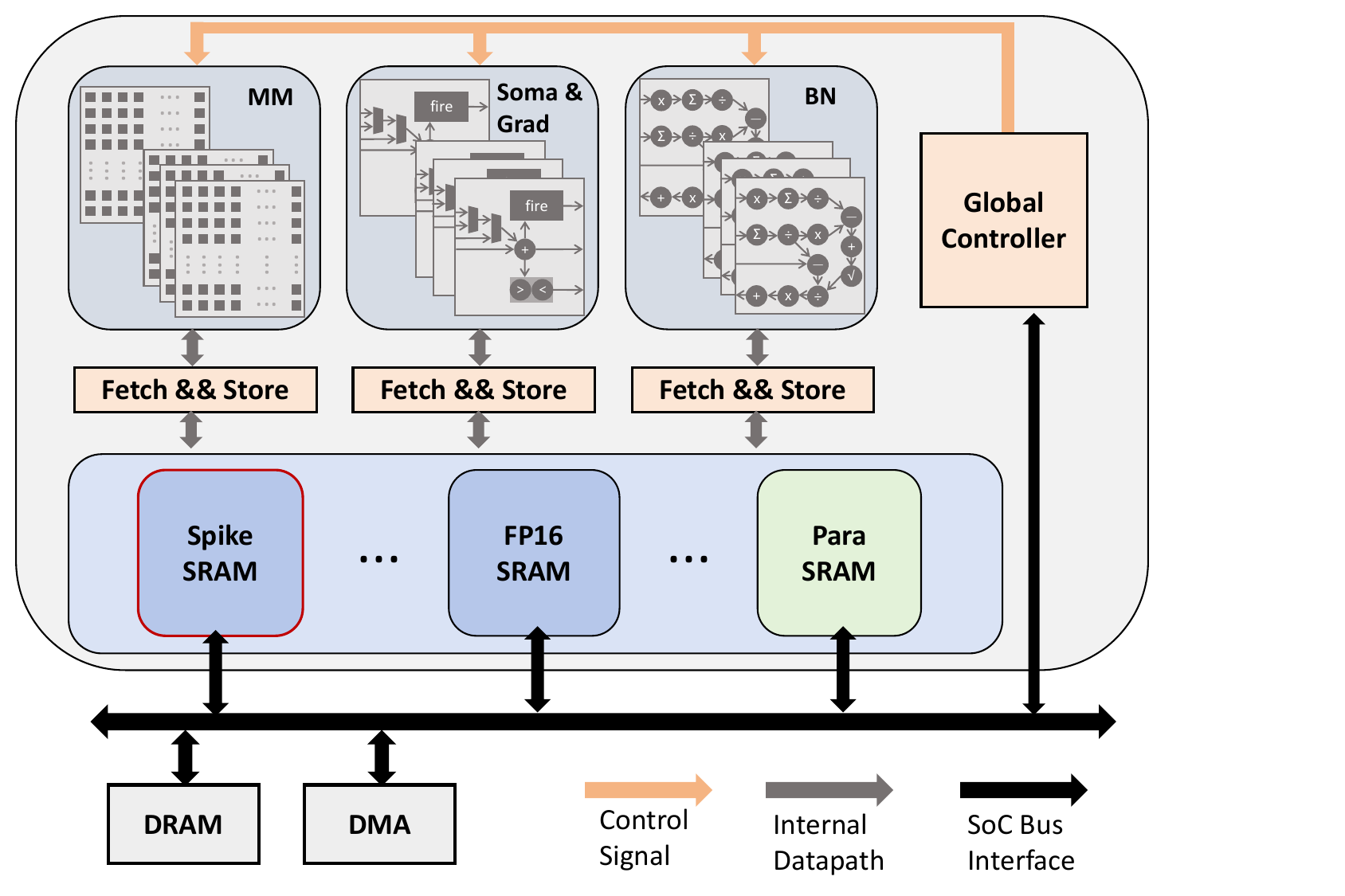}
\caption{The overall training architecture for Spiking Transformer.}
\label{fig_overall_archi}
\end{figure}

Fig. \ref{fig_overall_archi} shows the overall architecture of the accelerator for the training task in Spiking Transformer. Based on the analysis of Fig. \ref{fig_overall_training}, the main operations can be divided into three main parts, including the Matrix Multiply (MM), the SOMA and GRAD operation, and the batch normalization (BN) operation. These modules are controlled through a global controller to retrieve and store data from the memory hierarchy. For the MM module, the entire array is set as $64 \times 64$, supporting the operation of spike and FP16 addition in FP and WG, and FP16 multiplication in BP.

\begin{figure}[!t]
\centering
\includegraphics[width=0.5\textwidth,page=3,trim={30bp 20bp 150bp 0bp}, clip]{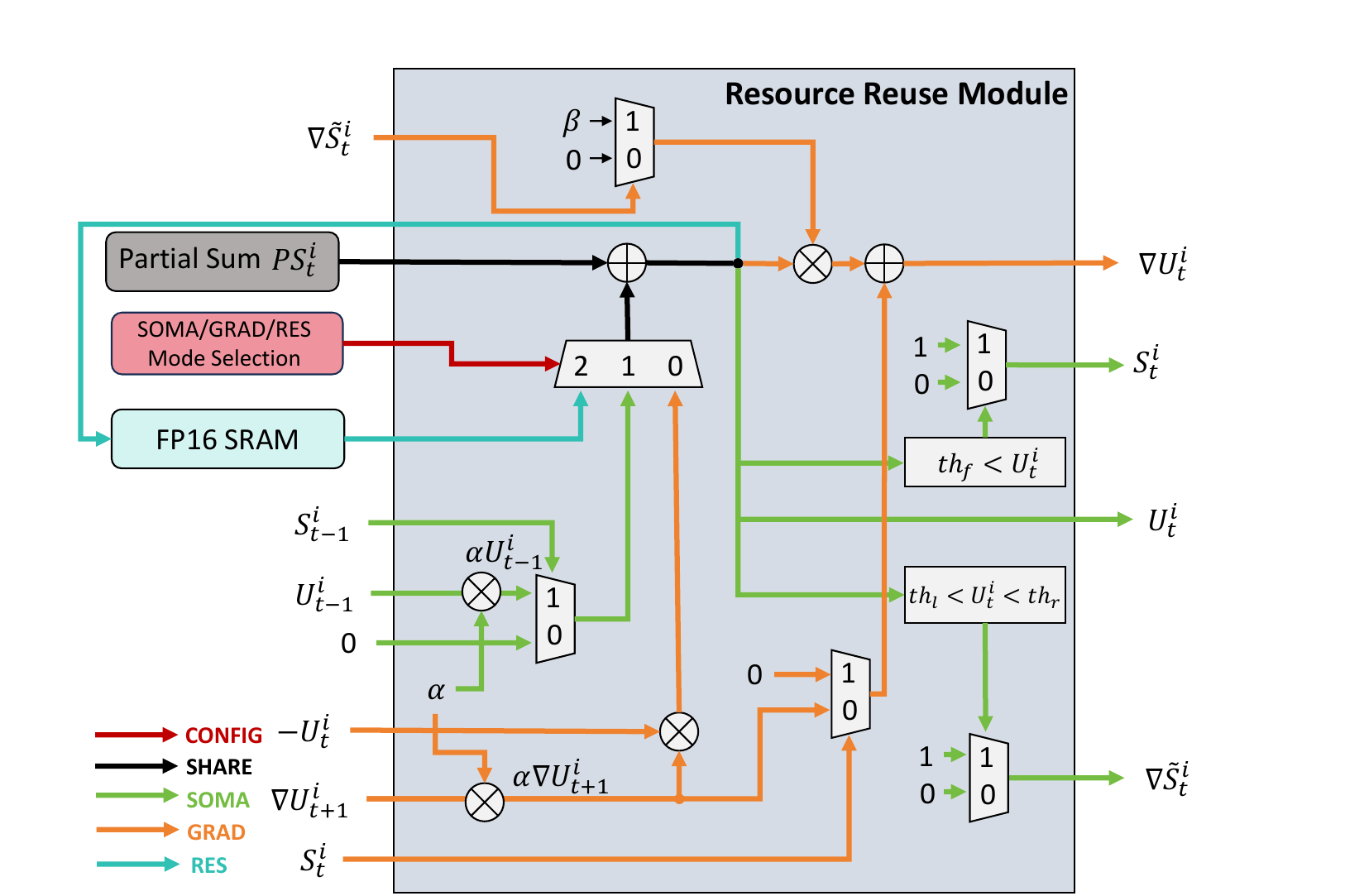}
\caption{The reuse framework for SOMA/GRAD/RES. The mode can be configured according to the selection.}
\label{fig_reuse}
\end{figure}

The SOMA, GRAD, and residual connection (RES) operations are integrated into a unified resource reuse framework, as depicted in Fig. \ref{fig_reuse}. This modular architecture supports three configurable operational modes: SOMA, GRAD, and RES, enabling dynamic switching based on the specific computational requirements of different training stages (FP, BP, WG). In SOMA mode, the module executes core spike-domain computations for FP. It takes as inputs the spike events from the previous time step, the corresponding membrane potentials, the decay parameter, and accumulated partial sums. Through a sequence of spike generation rules and membrane potential update mechanisms (consistent with the algorithmic details outlined earlier), it produces three key outputs: the updated membrane potential for the current time step, the newly generated spike events, and the spike gradient mask that encodes spatial-temporal spike patterns. When operating in RES mode, the module activates the residual connection pathway, primarily performing element-wise matrix addition to fuse feature maps. As indicated by the cyan data flow in Fig. \ref{fig_reuse}, this mode routes intermediate results to accumulate partial sums, preserving gradient flow integrity and mitigating vanishing gradients in deep spike networks. For GRAD mode, which implements the BP counterpart of SOMA, the module focuses on gradient computation for membrane potentials. It accepts inputs including the current membrane potential, spike events, the product of parameter \(\alpha\) and gradients of the previous membrane potential, and the spike gradient mask. Through chain-rule-based differentiation of spike activation functions and membrane dynamics, it generates the gradient of the current membrane potential, critical for weight update in the training loop. This integrated design enables efficient resource reuse across distinct computational phases, minimizing hardware overhead while ensuring seamless transition between forward and backward operations in the training.

\begin{figure}[!t]
\centering
\includegraphics[width=0.5\textwidth,page=4,trim={10bp 250bp 100bp 0bp}, clip]{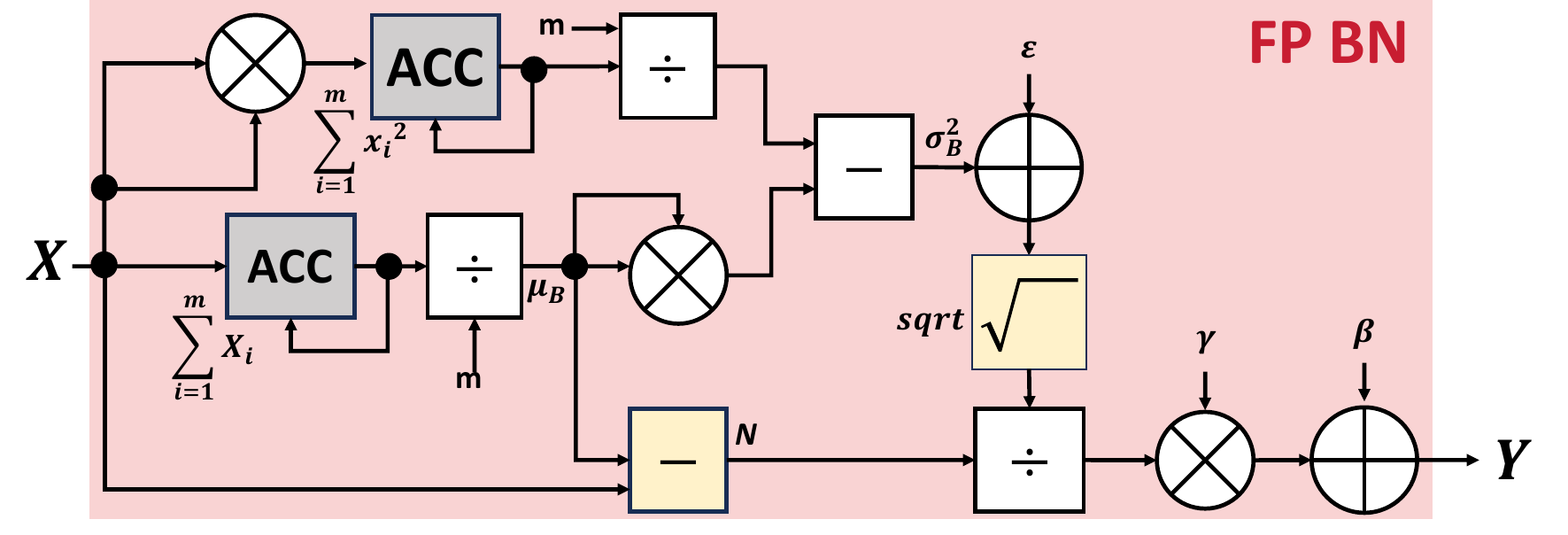}
\caption{The BN in FP process, following the algorithm description.}
\label{fig_FPBN}
\end{figure}

Fig. \ref{fig_FPBN} illustrates the fundamental architecture for the FP computation of the BN operation. Specifically, each sample processing path within this architecture integrates four adders, three multipliers, two dividers, two subtractors, and one square root unit, all optimized to support the FP16 data format, consistent with the precision requirements of our SpikingFormer-like Spiking Transformer training task. This component-level configuration enables the complete execution of forward BN operations, including mean calculation, variance computation, normalization, and scaling/shift steps, in a streamlined hardware pipeline, which is strictly aligned with the algorithmic workflow.
For a batch of $m$ samples with $D$ features (\(x_{i,d}\) denotes the $d$-th feature of the $i$-th sample), the FP BN computes:\begin{equation}
\mu_{B,d} = \frac{1}{m} \sum_{i=1}^m x_{i,d}
\end{equation}\begin{equation}
\bar{x^2}_d = \frac{1}{m} \sum{i=1}^m x_{i,d}^2
\end{equation}\begin{equation}
\sigma^2_{B,d} = \bar{x^2}_d - \mu_{B,d}^2
\end{equation}\begin{equation}
\mathit{sqrt}_d = \sqrt{\sigma^2_{B,d} + \varepsilon}
\end{equation}\begin{equation}
N_{i,d} = x_{i,d} - \mu_{B,d}
\end{equation}\begin{equation}
y_{i,d} = \gamma_d \cdot \frac{N_{i,d}}{\mathit{sqrt}_d} + \beta_d
\end{equation}
During training, \(\boldsymbol{\mu}_{B,d}\) and \(\boldsymbol{\sigma}_{B,d}^2\) are used directly, while inference employs running averages (updated via momentum) to ensure consistent hardware behavior across batch sizes, with statistics stored in on-chip memory for low-latency access.

\begin{figure}[!t]
\centering
\includegraphics[width=0.5\textwidth,page=5,trim={10bp 0bp 85bp 0bp}, clip]{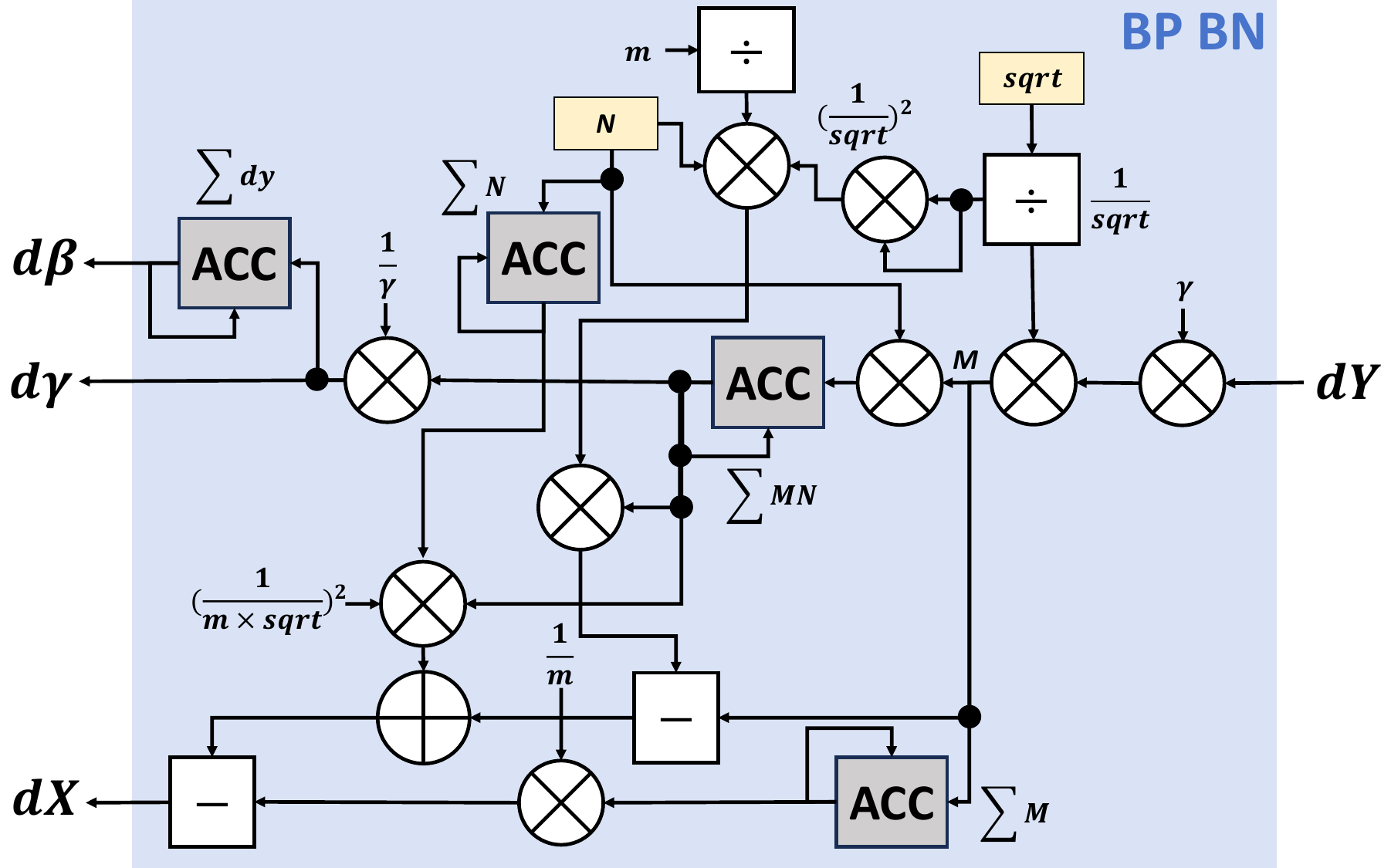}
\caption{The BN in BP process, following the algorithm description.}
\label{fig_BPBN}
\end{figure}

Correspondingly, Fig. \ref{fig_BPBN} depicts the hardware architecture tailored for the BP phase of the BN operation, following the algorithm in  \cite{BPBN}.
Given the gradient \(g_{i,d} = \frac{\partial \mathcal{L}}{\partial y_{i,d}}\) from subsequent layers, the gradients for BN parameters and inputs are derived as:\begin{equation}
M_{i,d} = \gamma_d \cdot \frac{g_{i,d}}{\mathit{sqrt}_d}
\end{equation}\begin{equation}
S_{N,d} = \sum_{i=1}^m N_{i,d},S_{M,d} = \sum_{i=1}^m M_{i,d},S_{MN,d} = \sum_{i=1}^m M_{i,d}N_{i,d}
\end{equation}\begin{equation}
\frac{\partial \mathcal{L}}{\partial \gamma_d} = \frac{1}{\gamma_d} \cdot S_{MN,d}
\end{equation}\begin{equation}
\frac{\partial \mathcal{L}}{\partial \beta_d} = \sum_{i=1}^m g_{i,d}
\end{equation}\begin{equation}
\frac{\partial \mathcal{L}}{\partial x_{i,d}} = M_{i,d} - N_{i,d} \cdot \frac{S_{MN,d}}{m \cdot \mathit{sqrt}_d^2} + \frac{S_{N,d} \cdot S_{MN,d}}{\mathit{sqrt}_d^2 \cdot m^2} - \frac{S_{M,d}}{m}
\end{equation}
The yellow boxes of the FP BN are delivered to BP BN. To meet the high-throughput demands of end-to-end training, all computational modules involved in BN are also deeply pipelined. This pipelining strategy overlaps the execution of consecutive BN operations across different batches or time steps, minimizing idle cycles and ensuring efficient utilization of arithmetic units even as the dataflow scales with batch size. The FP architecture retains compatibility with FP16 precision as well, maintaining consistency with the overall training precision scheme while supporting gradient computations for learnable BN parameters (scaling factor and shift term) and input features. ACC in both figures stands for the accumulation.

\section{Performance Simulation}
\subsection{Overview}
The proposed architecture is emulated via a synergistic integration of several components: our spike property simulator and the matrix computing framework with a latency modeling framework \cite{latency_ref}. Specifically, our spike property simulator is designed to meticulously track and accumulate the energy consumption and latency associated with computation and memory access throughout the entire training process. It focuses on critical spike-domain entities, including spike events, membrane potential dynamics, and spike gradient masks, while also accounting for essential parameters required for BN operations. Complementing this, the ZigZag framework  \cite{9360462}, which lacks the SNN property simulation, contributes by modeling matrix multiplication operations at both the computational and memory hierarchy levels, enabling detailed analysis of data flow and resource utilization in dense linear algebra computations. Further, the latency modeling framework systematically analyzes how architectural choices and dataflow patterns impact the overall training latency in our study.

\begin{figure*}[!htbp]
    \centering
    \includegraphics[width=1\textwidth,page=6,trim={10bp 50bp 200bp 100bp}, clip]{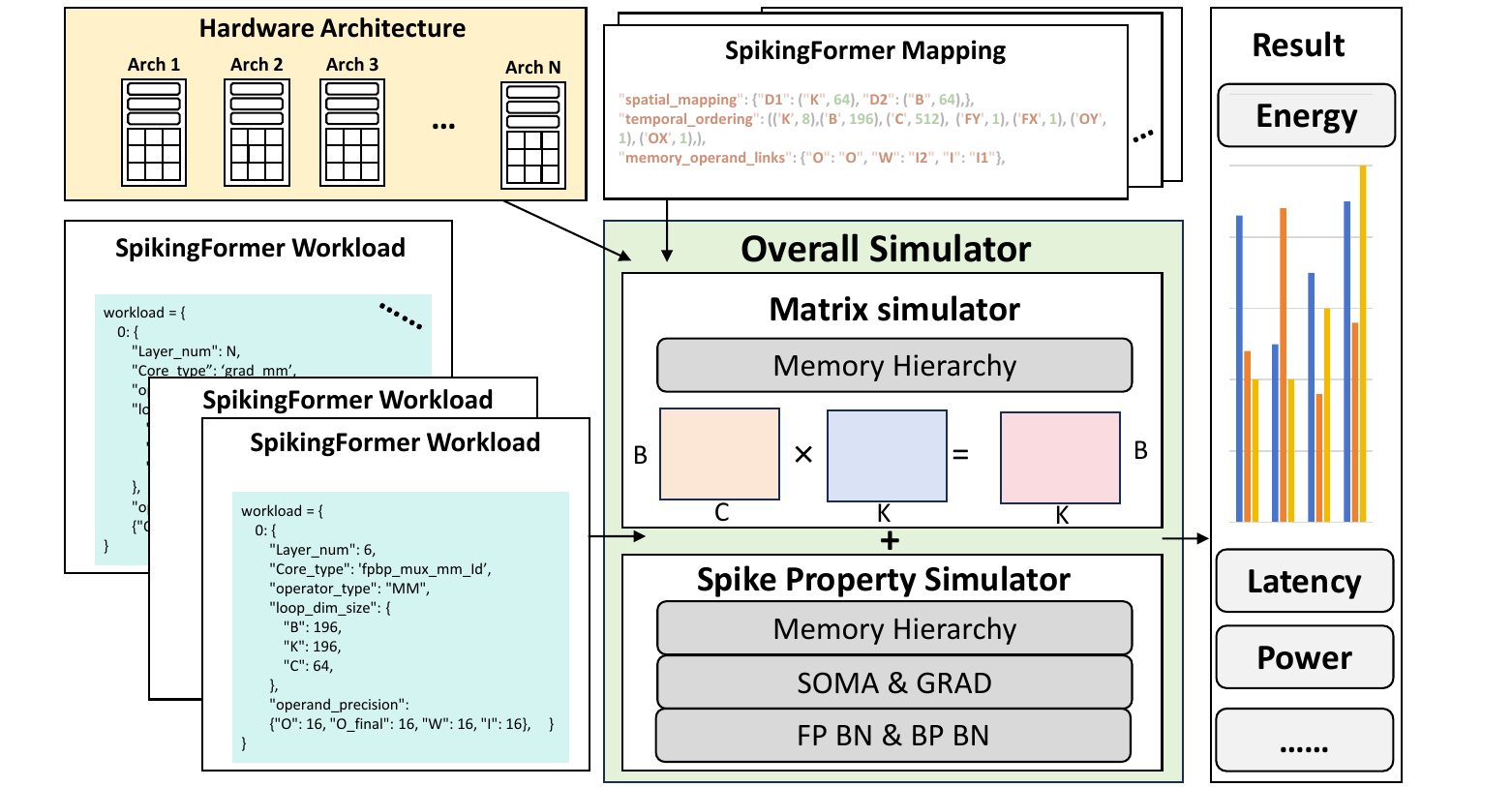}
    \caption{The overall simulation for the architecture by using the ZigZag \cite{9360462} simulation framework for the MM simulation. The simulation of SOMA and GRAD as well as the FP/BP BN are also achieved.}
    \label{fig_zigzag}
\end{figure*}

As illustrated in Fig. \ref{fig_zigzag}, the integrated training simulation framework for the accelerator operates as follows: leveraging ZigZag's configurable infrastructure, the simulator takes three key inputs: spike-driven Transformer workloads, diverse hardware architecture specifications, and a range of mapping strategies. The workloads encapsulate comprehensive details of the computations across the FP, BP, and WG stages, including operation types, tensor dimensions, and operand precision requirements. Notably, in the FP and WG stages, the input operand (I) is consistently represented as binary spikes with 1-bit precision, whereas the input operand in BP is encoded in FP16. Throughout all training phases, the weight (W) and output (O) operands maintain FP16 precision.

The target accelerator architecture features a fixed 64×64 processing array, interconnected with dedicated Static Random Access Memory (SRAM) modules for input, weight, and output storage. For workloads with varying matrix dimensions, the mapping behavior adapts dynamically based on a combination of spatial and temporal mapping strategies. In the FP stage, for instance, the input matrix in our experimental setup is defined with dimensions (B, C), while the weight matrix is configured as (C, K); this results in an output matrix of dimensions (B, K), with the mapping strategy optimizing data partitioning and scheduling to align with the 64×64 array constraints.

The output of the system shows the metrics of the accelerator in the overall training process, not only including the energy breakdown of stages in FP, BP and WG, but also containing the latency breakdown of FP, BP and WG. Moreover, due to various spatial mapping methods, different dataflows are generated with different latencies and energy amounts, which will be discussed in the following section.

\begin{table}[!t]
    \centering
    \caption{Description of Key Parameters}  
    \label{Description_symbols}  
    \renewcommand{\arraystretch}{1.3}  
    \begin{tabular}{c p{6cm}}  
        \toprule  
        Parameters & Description \\
        \midrule  
        $h$ & Head number (set to 8 in our work) \\
        $P$ & Patch size (set to 14 in our work) \\
        $d_{model}$ & Feature dimension (set to 512 in our work) \\
        $d_h$ & Feature dimension per head, calculated as $d_h = d_{model}/h$ \\
        $S$ & Sequence length, calculated as $S = BS \times T \times P^2$ \\
        $b$ & Maximum bitwidth (set to 16 in our work) \\
        $T$ & Timestep (set to 4 in our work) \\
        $BS$ & Batch Size (set to 16 in our work) \\
        $s_s$ & Spike sparsity \\
        $s_{smg}$ & Sparsity of spike gradient mask gradient \\
        $s_{pg}$ & Sparsity of potential gradient \\
        \bottomrule  
    \end{tabular}
\end{table}

\subsection{Detailed Simulation for FP, BP and WG}

\begin{table}[!t]
    \centering
    \caption{The computation energy of each operator in FP.}
    \label{energy_com_FP}  
    \includegraphics[width=0.5\textwidth, page=13, trim={0bp 140bp 350bp 0bp}, clip]{TCAD.pdf}
\end{table}

Similar to SATA \cite{SATA}, we have developed a series of energy evaluation methods in this study. The key difference lies in that we propose detailed energy consumption metrics for memory access and computation corresponding to different operations throughout the training task, including the FP, BP, and WG stages. Table \ref{Description_symbols} presents the parameters used along with their respective meanings or origins, and these parameters are frequently referenced in the subsequent tables. Generally, the energy is shown as follows:
\begin{equation}
E^{Total}_{X} = E^{FP,Total}_{X}+E^{BP,Total}_{X}+E^{WG,Total}_{X}
\end{equation}\
The $Total$ indicates the combination of $C$ and $M$, with $C$ for computation energy and $M$ for memory access energy. The $X$ denotes the operations among MM, BN, LIF and RES.
\begin{equation}
E^{Total} = E^{C}+E^{M}
\end{equation}\

Based on the training workflow illustrated in Fig. \ref{fig_overall_training}, the FP energy evaluation (marked in red) is systematically decomposed into two core components: computation energy, which is elaborated in Table \ref{energy_com_FP} with quantitative breakdowns, and memory access energy, as explicitly presented in Table \ref{energy_mem_FP}. By leveraging a well-validated existing process library (28nm in our work), we conduct a fine-grained accounting of the basic computation energy corresponding to element-wise operations in FP, encompassing fundamental arithmetic and logic primitives such as $E_{MAC}$, $E_{ADD}$, $E_{SUB}$, $E_{MUL}$, $E_{MUX}$, $E_{SQRT}$, and $E_{DIV}$. Notably, sparsity effects, an inherent characteristic of spike-driven computations, are rigorously incorporated into these energy calculations to ensure alignment with real-world execution patterns. In the context of the tables, the notation $l$ specifically refers to the layer index within the Spike Transformer Block, enabling layer-specific energy profiling.

\begin{table}[!t]
    \centering
    \caption{The memory access energy of each operator in FP.}
    \label{energy_mem_FP}
    \includegraphics[width=0.5\textwidth, page=17, trim={0bp 240bp 300bp 0bp}, clip]{TCAD.pdf}
\end{table}

The energy estimation for MM and RES computations exhibits a direct and interpretable relationship with matrix dimensions, as their computational intensity scales predictably with factors such as input/output channel size and spatial dimensions. In contrast, the computation energy for SOMA and BN operations is more complex, as it depends on architecture-specific parameters (e.g., membrane potential decay coefficients in SOMA) and algorithmic details (e.g., batch statistics updating rules in BN) outlined in the preceding sections, requiring a tailored modeling approach. For memory access energy, we comprehensively analyze storage access behaviors associated with each operation, spanning the entire three-level memory hierarchy from off-chip Dynamic Random Access Memory (DRAM) to on-chip SRAM and register files. Critical parameters, including data bitwidth, access frequency, and read/write energy per transaction, are explicitly quantified and summarized in the Table \ref{rdwr} to support accurate energy accounting. In system-on-chip (SoC) designs, DRAM access incurs significant power consumption and often constitutes a dominant portion of total energy usage \cite{sramdram}. Therefore, while maintaining a reasonable on-chip SRAM capacity, the number of read and write operations between DRAM and SRAM should be minimized, with as many memory accesses 
as possible handled solely within SRAM. Additionally, every register has one latency to simulate the entire latency.

It is meaningless that the results of BN in FP and square root values are retained in intermediate storage to ensure the integrity of subsequent BP BN propagation. At the same time, membrane potential and spike timing information generated by the SOMA module during FP are persistently stored to provide essential input for GRAD computations in the WG stage.  

\begin{table}[!t]
    \centering
    \caption{Memory Parameters and Energy Consumption}
    \label{rdwr}
    \renewcommand{\arraystretch}{1.3} 
    \begin{tabular}{c c c c c} 
        \toprule
        Number & Memory Type & Memory Volume & Bitwidth & Energy (pJ/bit) \\
        \cmidrule(r){1-1} \cmidrule(lr){2-2} \cmidrule(lr){3-3} \cmidrule(lr){4-4} \cmidrule(l){5-5}
        & & & & Read / Write \\
        \midrule
        0 & DRAM & - & - & $d_0^r$ / $d_0^w$ \\
        1 & SRAM & $V_1$ & 1bit & $s_0^r$ / $s_0^w$ \\
        2 & SRAM & $V_2$ & 16bit & $s_1^r$ / $s_1^w$ \\
        3 & SRAM & $V_3$ & 16bit & $s_2^r$ / $s_2^w$ \\
        4 & SRAM & $V_4$ & 16bit & $s_3^r$ / $s_3^w$ \\
        5 & SRAM & $V_5$ & 16bit & $s_4^r$ / $s_4^w$ \\
        6 & SRAM & $V_6$ & 16bit & $s_5^r$ / $s_5^w$ \\
        7 & SRAM & $V_7$ & 1bit & $s_6^r$ / $s_6^w$ \\
        8 & SRAM & $V_8$ & 16bit & $s_7^r$ / $s_7^w$ \\
        9 & SRAM & $V_9$ & 16bit & $s_8^r$ / $s_8^w$ \\
        9 & Register & - & 1bit & $r_0^r$ / $r_0^w$ \\
        10 & Register & - & 16bit & $r_1^r$ / $r_1^w$ \\
        \bottomrule
    \end{tabular}
\end{table}

\begin{table}[!t]
    \centering
    \caption{The computation energy of each operator in BP and WG.}
    \label{energy_com_BPWG}  
    \includegraphics[width=0.5\textwidth, page=14, trim={0bp 40bp 0bp 0bp}, clip]{TCAD.pdf}
\end{table}

The BP and WG's computation and memory access energy are detailed in Table \ref{energy_com_BPWG} and Table \ref{energy_mem_BPWG}, respectively. Consistent with FP, sparsity effects are integrated into the computation energy models for BP and WG, reflecting the sparse activation patterns inherent in spike-driven training. MM, GRAD, and RES operations among the computation components exhibit relatively straightforward energy characteristics due to their regular arithmetic patterns. In contrast, BN operations in BP and WG involve more intricate energy modeling. As dictated by the complex algorithmic flow described earlier, BN operations encompass up to eight distinct sub-components (e.g., gradient computation for batch mean and variance), each requiring independent energy evaluation to ensure the accuracy of the overall training energy profile. Regarding memory access, the energy modeling framework for BP and WG follows the same three-level hierarchy and parameterization strategy as FP, ensuring consistency and comparability across all training stages.
\begin{table}[!t]
    \centering
    \caption{The memory access energy of each operator in BP and WG.}
    \label{energy_mem_BPWG}  
    \includegraphics[width=0.5\textwidth, page=15, trim={0bp 170bp 50bp 0bp}, clip]{TCAD.pdf}
\end{table}

\section{Experiment}
\subsection{Experiment Setup}
To validate the practical efficacy of this design and assess it under reasonable hardware resource and computational complexity constraints, 
this study conducts experiments on the representative ImageNet-1K dataset, thereby avoiding the scale and complexity limitations of smaller datasets. The configuration of the Spiking Transformer model is settled in Table \ref{Description_symbols}. The memory sizes are set to contain both the FP16 signals and the spike binary signals, and are modeled as different memories with derived energy using CACTI \cite{cacti}.
\subsection{Experimental Analysis}
In experiments, due to the extensive computational processes of the Spiking Transformer, it is partitioned into multiple layers to align with the concept of different layers in the workload. 

In FP, the spatial unfolding scenario with "D1": B and "D2": C in the spatial mapping is regarded as input stationary (IS), as these two dimensions are related to the input matrix. Here, "D1" and "D2" have the meaning of spatial mapping dimensions. Similarly, the spatial unfolding scenario with "D1": C and "D2": K is considered weight stationary (WS), since these dimensions are associated with the weights. The spatial unfolding with "D1": B and "D2": K corresponds to output stationary (OS). 

Notably, in the BP process, for the IS mode, the dimensions become "D1": B and "D2": K, because the input dimension at this point is the output dimension of the forward propagation. For the WS mode, the spatial dimensions remain consistent with those in the forward propagation. For the OS mode, the dimensions need to be adjusted to "D1": B and "D2": C, as the output in BP corresponds to the input activation values of the forward propagation. 

\begin{figure}[!t]
\centering
\includegraphics[width=0.5\textwidth,page=7,trim={10bp 0bp 130bp 0bp}, clip]{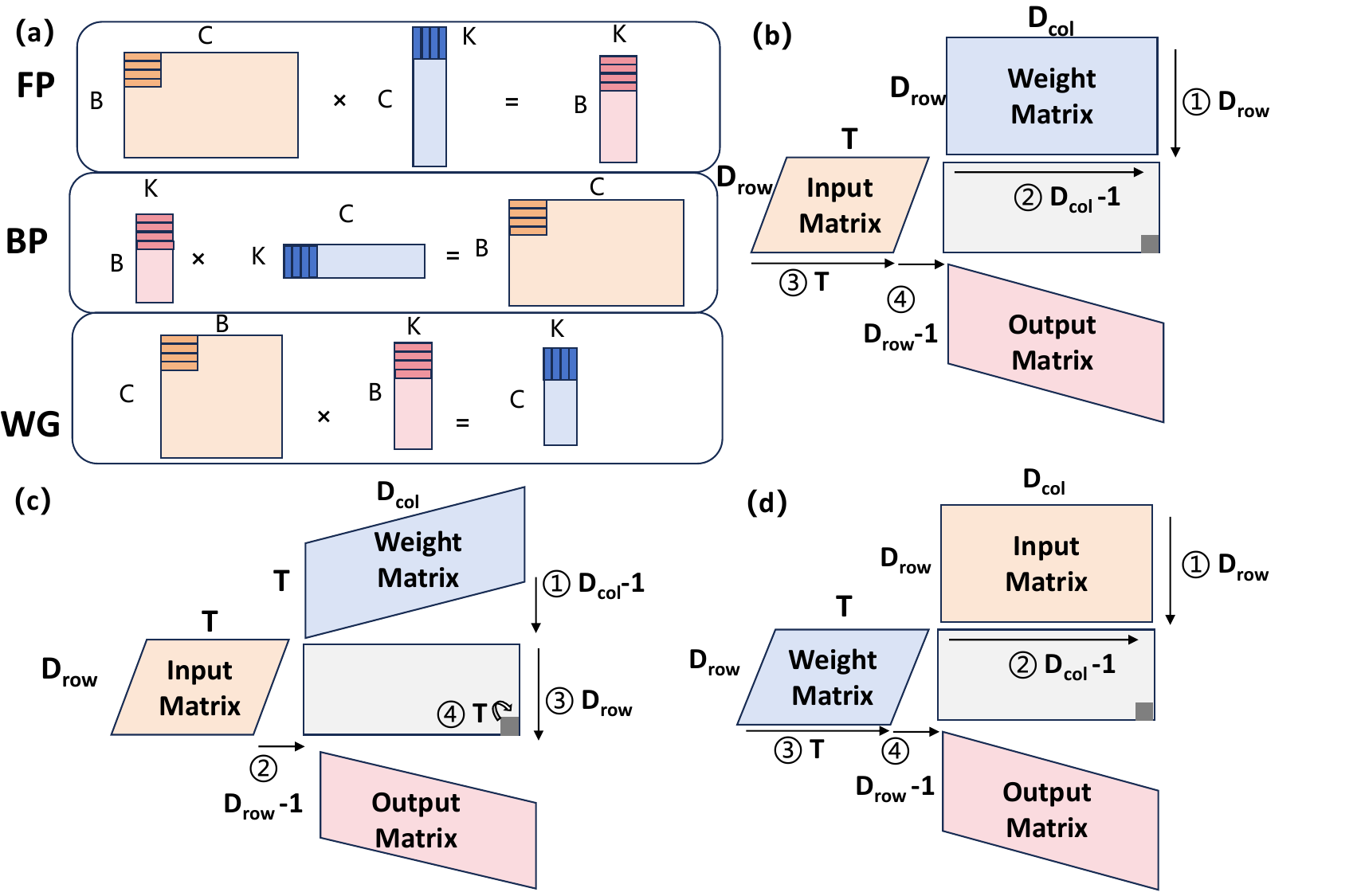}
\caption{The systolic array computation under given dimensions: input (B, C), weight (C, K), output (B, K). (a) The dimension changes in FP, BP, and WG. (b) The weight stationary mode. (c) The output stationary mode. (d) The input stationary mode.}
\label{fig_systolic}
\end{figure}

Similarly, in the WG process, the input stationary (IS) corresponds to "D1": B and "D2": C, the WS corresponds to "D1": B and "D2": K, and the OS corresponds to "D1": C and "D2": K, as illustrated in Fig. \ref{fig_systolic}. These modes are mentioned in \cite{scale1} and \cite{scale2}. For the matrix multiplication latency, considering various dimensions of different dataflow formats as shown in the Fig. \ref{fig_systolic}, the runtime of the dataflows can be summarized by the following equation, which takes into account the time consumed by input distribution, partial-sum accumulation, and result transmission:
\begin{equation}
\label{systolic_t}
t = 2D_{row} + D_{col} + T - 2
\end{equation}
When the matrix dimensions (e.g., \(B \times C\) multiplied by \(C \times K\)) are larger than the array dimensions, the matrix needs to be decomposed into multiple sub-blocks. Consequently, the above formula is modified as:
\begin{equation}
\label{systolic_T}
t_{\text{total}} = (2D_{row} + D_{col} + T - 2) \times \left\lceil \frac{B}{D_{row}} \right\rceil \times \left\lceil \frac{K}{D_{col}} \right\rceil
\end{equation}
Accordingly, the array utilization efficiency can be expressed as:
\begin{equation}
\label{utilization}
\eta = \frac{B \times C \times K}{t_{\text{total}} \times D_{row} \times D_{col}}
\end{equation}
These methods are critical in utilization calculation of systolic-like accelerators  \cite{COSA}, \cite{Transformer_train}, \cite{LLMCompass}, \cite{SIGMA}, \cite{34}.

\begin{figure}[!t]
    \centering
    \includegraphics[width=0.5\textwidth,page=8,trim={0bp 40bp 0bp 20bp}, clip]{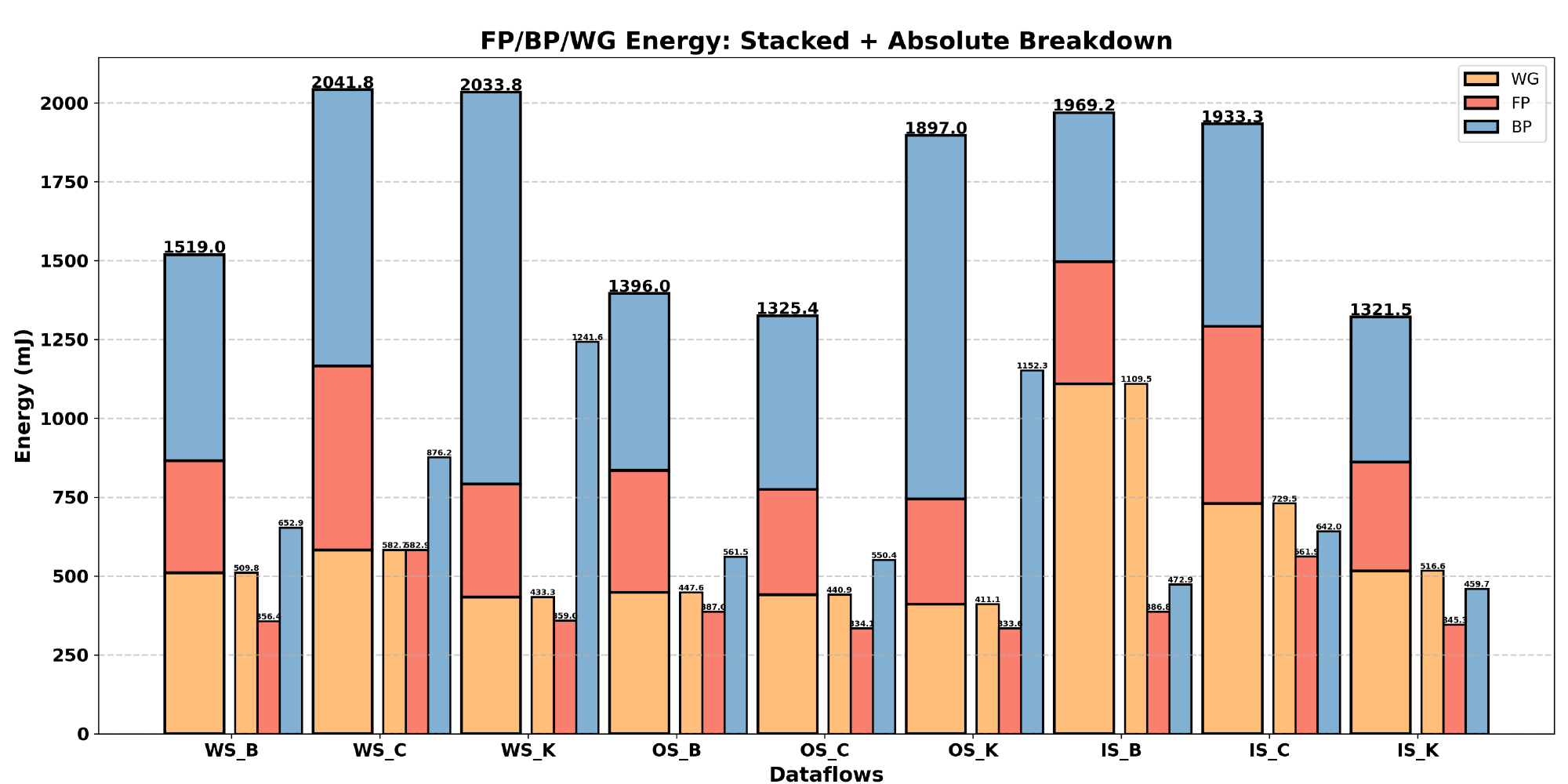}
    \caption{The  energy breakdown in FP, BP and WG of nine dataflows.}
    \label{fig_overall_energy}
\end{figure}
\begin{figure}[!t]
\centering
\includegraphics[width=0.5\textwidth,page=9,trim={40bp 10bp 20bp 0bp}, clip]{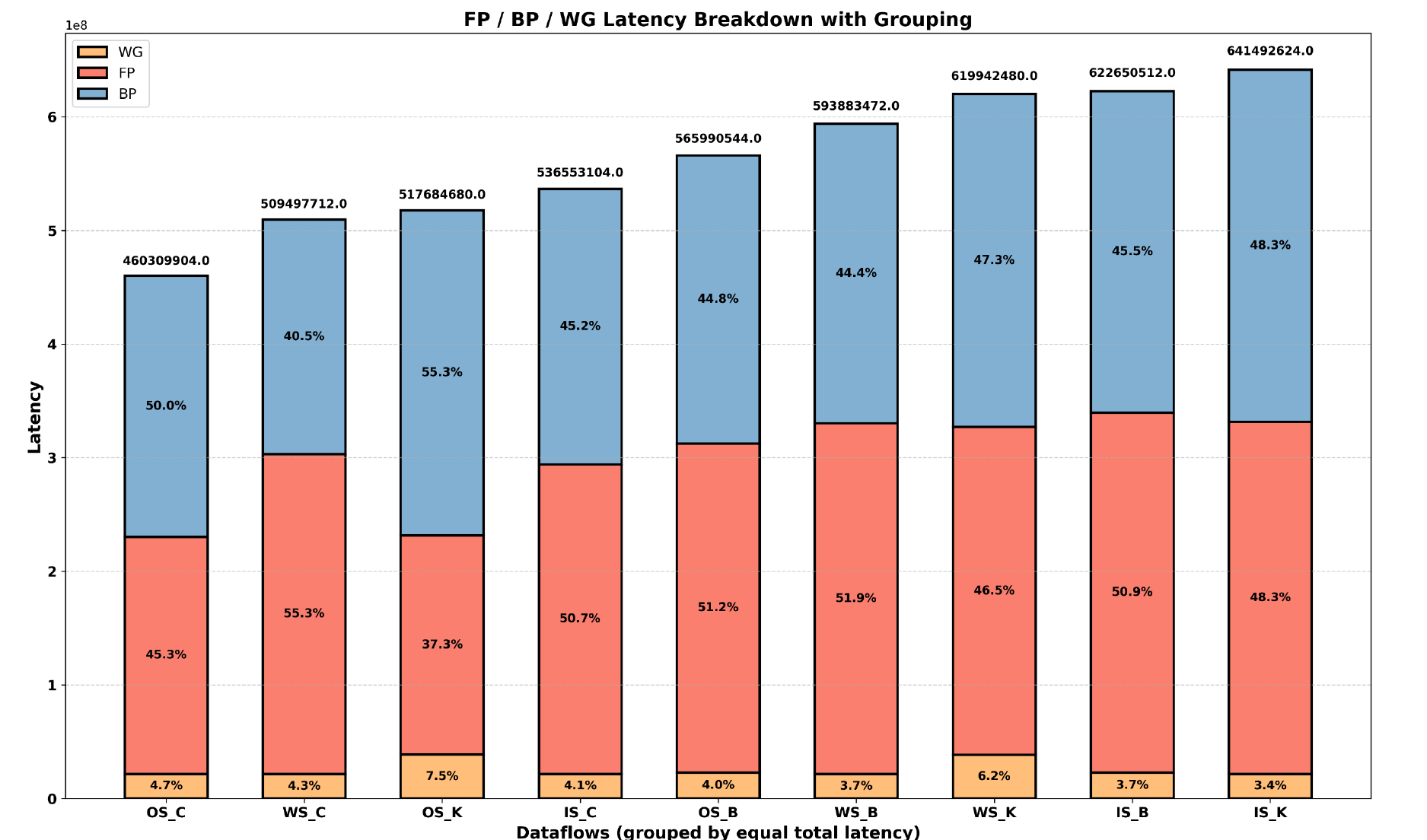}
\caption{The overall latency breakdown.}
\label{fig_overall_latency}
\end{figure}
\subsection{Experimental Results}
Furthermore, nine dataflow schemes are derived through our experiments, combining three internal dataflow modes with external matrix partitioning across three dimensions. Each dataflow scheme is configured to simulate the computational processes of FP, BP, and WG respectively, yielding their respective energy consumption simulation breakdowns and latency simulation breakdowns.
\begin{figure*}[!htbp]
    \centering
    \includegraphics[width=1\textwidth,page=10,trim={0bp 170bp 50bp 60bp}, clip]{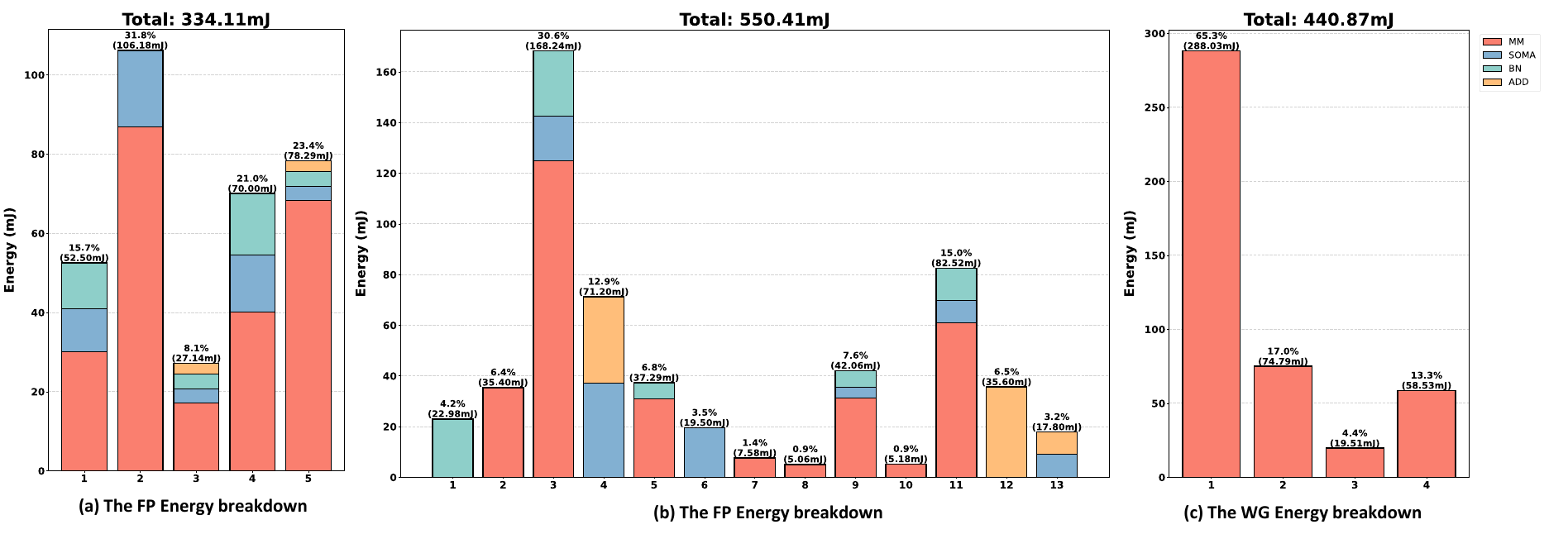}
    \caption{Energy breakdown in (a) FP (b) BP and (c) WG for OS\_C dataflow.}
    \label{fig_overall_energy_breakdown}
\end{figure*}
In terms of energy shown in Fig. \ref{fig_overall_energy}, overall, due to the complexity of BP, it accounts for a significant proportion of the total energy, nearly exceeding the energy consumption of both FP and WG. Among these schemes, the dataflow combination with the lowest total energy is OS\_C, which means that internally, it adopts the OS dataflow as described earlier. Externally, the matrix partitioning follows the C dimension in the looping manner, where the C dimension here, exemplified by the C dimension in FP, refers to the column dimension of the input matrix and the row dimension of the weight matrix.

In terms of latency shown in Fig. \ref{fig_overall_latency}, similarly, it presents the latency breakdown of FP, BP, and WG across dataflows grouped by equal total latency. The OS\_C configuration achieves the lowest cumulative latency, delivering a notable reduction of approximately 10\%–28\% compared to other dataflows.

In more detail, we break down the energy consumption by operator for the FP, BP, and WG phases of the dataflow OS\_C with the lowest energy, which is consistent with our previous work SimST [need cite] for Spiking Transformer inference. In each subfigure in Fig. \ref{fig_overall_energy_breakdown}, the red portion represents MM, the blue portion corresponds to SOMA (SOMA in FP, GRAD in BP), the green portion denotes BN, and the yellow portion stands for residual connections' addition. MM accounts for most of the figures, indicating that matrix multiplication involves extensive memory access and remains the task's most critical and largest component. The entire FP phase is divided into five stages, corresponding to Fig. \ref{fig_overall_training}: the first stage corresponds to the Q/K/V linear transformation layer, the second stage to SSA, the third stage to the Z linear transformation layer, the fourth stage to the A linear transformation layer, and the fifth stage to the B linear transformation layer. Besides, WG is also divided into four stages, namely \(W_B\), \(W_A\), \(W_O\), and \(W_{Q/K/V}\) updates. Due to its relatively high computational complexity, BP is divided into 13 stages, as shown in the following Fig. \ref{fig_bp_stages}. Unlike FP, BP's V matrix must be obtained first to calculate the attention score gradient.

\begin{figure}[!t]
\centering
\includegraphics[width=0.5\textwidth,page=11,trim={0bp 50bp 0bp 100bp}, clip]{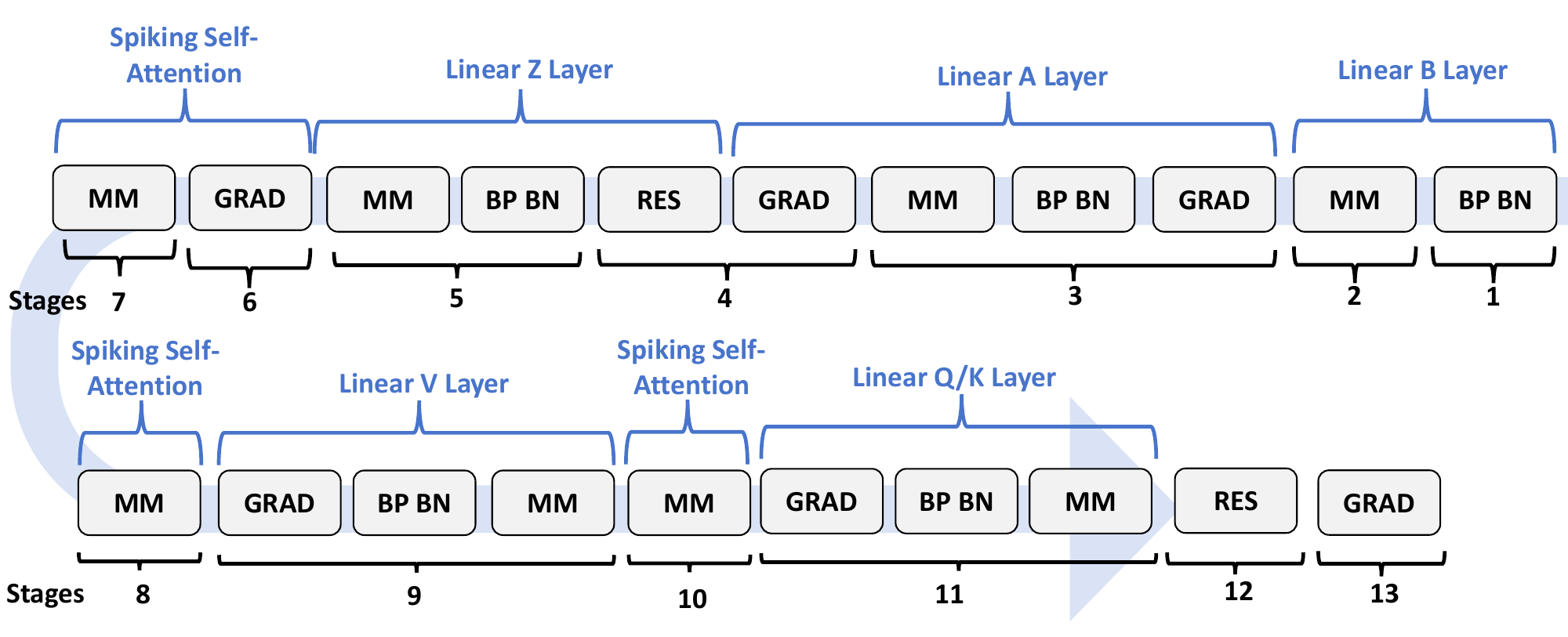}
\caption{The detailed stages in BP.}
\label{fig_bp_stages}
\end{figure}
\subsection{Comparison and Discussion}
\begin{table*}[!t]
    \renewcommand{\arraystretch}{1.6} 
    \setlength{\tabcolsep}{5pt}
    \caption{Comparison of Key Metrics Across Related Works}
    \label{tab:comparison}
    \centering
    \footnotesize
    \begin{tabular}{m{2cm}*{10}{m{1.2cm}}}
        \hline
        \textbf{Items} & \textbf{This Work} & \textbf{SIGMA \cite{SIGMA}} & \textbf{SVLSI'20 \cite{34}} & \textbf{H2Learn \cite{H2Learn}} & \textbf{SATA \cite{SATA}} & \textbf{COSA \cite{COSA}} & \textbf{TRETA \cite{Transformer_train}} & \textbf{ArXiv'25 \cite{NC}} & \textbf{TPU-like Engine \cite{google_tpu_whitepaper}} & \textbf{GPU V100 \cite{nvidia_v100_whitepaper}} \\
        \hline
        Process (nm) & 28 & 28 & 14 & 28 & 65 & - & 28 & 28 & 28 & 12 \\
        \hline
        Frequency (MHz) & 500 & 500 & 1000 & 800 & 400 & 200 & 500 & 500 & 500 & 1530 \\
        \hline
        Model Support & Spiking Transformer & DNN & CNN/RNN & SNN & SNN & Transformer & Transformer & SNN & DNN/SNN & DNN/SNN \\
        \hline
        Precision & FP16 & FP16 & FP16 & FP16 & FP16 & INT8 & PINT(8,3) & FP16 & FP16 & FP16 \\
        \hline
        MAC Configuration & 64×64 & 128×128 & 2×8×8×8 & 64×64 & 512 & 64×64 & 8×8×16×16 & 32×3×16×16 & 128×128 & 640×64 \\
        \hline
        Efficient Throughput & 3.4 TFLOPS & 10.8 TFLOPS & 2 TFLOPS & 27.85 TFLOPS & - & 1556.5 GOPS/s & 14.71 TOPS & 16 TFLOPS & 1.88 TFLOPS & 15.7 TFLOPS \\
        \hline
        Power (W) & 1.44 & 22.33 & 1.43 & 20.57 & - & 31.3 & 4.45 & 14.49 & 12.25 & 300 \\
        \hline
        Energy Efficiency & 2.36 TFLOPS/W & 0.48 TFLOPS/W & 1.4 TFLOPS/W & 1.354 TFLOPS/W & - & 49.7 GOPS/J & 3.31 TOPS/W & 1.05 TFLOPS/W & 0.15 TFLOPS/W & 0.053 TFLOPS/W \\
        \hline
        Training Support & Yes & Yes & Yes & Yes & Yes & No & Yes & Yes & Yes & Yes \\
        \hline
    \end{tabular}
\end{table*}
Based on the emulated energy consumption and latency metrics derived from our full-system simulation framework, we identified the optimal dataflow for the proposed architecture, with key performance benchmarks and comparative results summarized in Table \ref{tab:comparison}. Notably, the simulated power is calculated through the generated energy and latency of OS\_C, dividing the entire energy by the overall latency. This comparative analysis validates our design's effectiveness and contextualizes its advantages within the broader landscape of neuromorphic and deep learning accelerators.

Compared with these State-Of-The-Art (SOTA) works, in terms of hardware fundamentals, this work adopts a 28 nm process node, consistent with most related works like \cite{SIGMA} and \cite{H2Learn}, and balancing manufacturing feasibility with performance. Its operating frequency is set to 500 MHz, which is moderate compared to 1000 MHz in \cite{34} and 1530 MHz for GPU V100 \cite{nvidia_v100_whitepaper}, avoiding excessive power consumption caused by ultra-high frequencies.

For model support, this work is tailored for SpikingFormer-like Spiking Transformer, filling the gap in dedicated architectures for spiking transformer training, distinct from general DNN/CNN-focused designs like \cite{COSA} and pure SNN accelerators such as \cite{SATA} and \cite{H2Learn}. It retains FP16 precision, ensuring computational accuracy while aligning with mainstream deep learning hardware standards.

In terms of computational performance, this work achieves an efficient throughput of 3.4 TFLOPS, with overall MAC array utilization 83\% from formula \ref{utilization}. Although lower than the 27.85 TFLOPS of \cite{H2Learn}, our work demonstrates significant advantages in energy efficiency: its energy efficiency reaches 2.36 TFLOPS/W, which is 4.9× higher than \cite{SIGMA}, 1.7× higher than \cite{34}, and far superior to traditional accelerators such as TPU-like systolic engine and GPU V100. Compared with \cite{Transformer_train}, they provide PINT format, resulting in TOPS throughput, in which our work's TFLOPS throughput is also competitive. This efficiency gain stems from the optimized Spiking Transformer computation and memory hierarchy design, making it more suitable for resource-constrained scenarios.

Notably, this work supports training functionality, consistent with most advanced designs, considering the learning ability of the neuromorphic features, the same as \cite{NC}. This training capability is critical for deploying Spiking Transformers, as it enables direct hardware-aware fine-tuning. Overall, the proposed architecture balances the model compatibility and energy efficiency, particularly excelling in energy-efficient training support for Spiking Transformers based on the detailed training workflow and simulation method, providing a valuable reference for future neuromorphic accelerator design.

\section{Consequence}
In conclusion, this work addresses the critical gap in energy-efficient training of Spiking Transformers by proposing the first specialized acceleration architecture tailored for SpikingFormer-like Spiking Transformer models. By comprehensively modeling temporal and spatial spike signals during training and enabling systematic latency and energy simulations, the design fills the void of dedicated hardware solutions for such hybrid models, unlike existing accelerators focused on DNN inference, pure SNNs, or general Transformers.
Comparative analysis validates its strengths compared with SOTAs. This balance of specialization for Spiking Transformers, energy efficiency, and training support demonstrates practical feasibility for resource-constrained scenarios. It provides a foundational framework for future neuromorphic accelerator design targeting hybrid Spiking Transformer systems.

\printbibliography

\end{document}